\documentclass{optica-article}

\journal{opticajournal} 

\articletype{Research Article}

\usepackage{lineno}

\usepackage{float}
\usepackage{subfigure}
\usepackage{xcolor}
\usepackage{graphicx}
\usepackage{dcolumn}
\usepackage{bm}
\usepackage{physics}
\usepackage{siunitx}
\usepackage{ulem}
\usepackage{mathrsfs}

\usepackage{xr-hyper}
\usepackage{hyperref}
\hypersetup{
  colorlinks   = true, 
  urlcolor     = blue, 
  linkcolor    = red, 
  citecolor   = red 
}

\usepackage{cleveref}
\newcommand{\gt}{g^{(2)}}
\newcommand{\Gt}{G^{(2)}}

\begin{document}

\title{Aberration sensing in fluorescence microscopy using quantum correlations}

\author{Shay Elmalem,\authormark{1,*} Shaurya Aarav,\authormark{2,*} Allard K. F. Hübler,\authormark{3} Hugo Defienne,\authormark{2,\dag} and Dan Oron\authormark{1,\ddag}}

\address{\authormark{1}Department of Molecular Chemistry and Materials Science, Weizmann Institute of Science, Rehovot 76100, Israel\\
\authormark{2}Sorbonne Université, CNRS, Institut des NanoSciences de Paris, INSP, F-75005 Paris, France\\
\authormark{3}Georg-August-Universität, 37073 Göttingen, Germany\\
\authormark{*}These authors contributed equally}

\email{\authormark{\dag}hugo.defienne@insp.jussieu.fr} 
\email{\authormark{\ddag}dan.oron@weizmann.ac.il}

\begin{abstract*} 
Widefield fluorescence microscopy is an essential biological imaging tool, yet its penetration depth is fundamentally restricted by tissue-induced aberrations. 
Correcting these requires precise characterization of the point spread function (PSF), but existing wavefront sensing techniques rely either on often-unavailable guide stars, or on computationally heavy indirect phase retrieval methods, that could struggle in low-contrast regions. 
Here, we show that the PSF can be mapped directly and non-invasively by exploiting quantum correlations between photons. 
Standard fluorophores inherently exhibit photon antibunching, effectively acting as sources of ``missing'' photon pairs. 
By capturing these correlation statistics across a wide field of view using a single-photon avalanche diode (SPAD) array imager, we extract the local PSF autocorrelation directly from the fluorescence emission itself. 
This approach bypasses the limitations of classical intensity measurements, requiring no iterative or complex numerical optimization. 
We demonstrate widefield direct PSF retrieval in a second or less, paving the way for real-time operation. 
By translating quantum optical concepts to conventional fluorescent probes, this technique provides a robust, direct pathway to deep-tissue adaptive optics in life sciences.

\end{abstract*}

\section{Introduction} \label{sec:01_intro}

Widefield fluorescence microscopy is one of the most broadly used imaging techniques in life sciences. Yet, it is often limited to relatively shallow depths, typically on the order of a few scattering lengths, due to the deterioration of the image quality by scattering in the sample. Scattering induces a spatially variant wavefront aberration which can be detrimental to image resolution and contrast. It is often difficult to compensate for aberrations across a large area via standard tools of adaptive optics. 
To enable such correction, the local point spread function (PSF) of the imaging system must first be characterized (see~\cite{Hampson2021} and references therein). This can be achieved either directly via a wavefront sensor, or indirectly by extracting it from the observed images.
Notably, this task becomes more difficult at depth, where increased scattering shrinks the ``isoplanatic patch'' - the spatial extent over which the PSF remains invariant.

In practice, direct characterization methods often rely on the detected light being spatially coherent, necessitating the use of bright point sources (``guide stars'') within the sample \cite{Azucena2010,Tao2012}. This requirement significantly limits their applicability in fluorescence imaging.
In contrast, indirect methods (such as pupil segmentation \cite{Ji2010} and phase diversity \cite{Johnson2024}) avoid exogenous contrast agents but often require the acquisition of multiple images and a computationally intensive numerical optimization procedure to retrieve the local PSF. 
Furthermore, their performance is strongly sample-dependent, often struggling to accurately characterize the PSF in regions with low fluorescence contrast.

A promising alternative for direct PSF characterization that does not require an external wavefront sensor is quantum microscopy \cite{Defienne2024}. 
It employs information from photon correlations rather than intensity, as correlations can encode additional information inaccessible to direct intensity measurements. 
Indeed, it was recently shown that adaptive optics can be performed in transmission microscopy when the illumination is composed of photon pairs generated by spontaneous parametric down conversion (SPDC), by optimizing on the correlation signal instead of the intensity \cite{Cameron2024}. 
While this label-free configuration seems to significantly differ from conventional fluorescence microscopy, these concepts can be translated to fluorescence imaging, provided an appropriate quantum resource can replace the entangled photon source.

\begin{figure*}[t]
\centering
\includegraphics[width=0.8\linewidth]{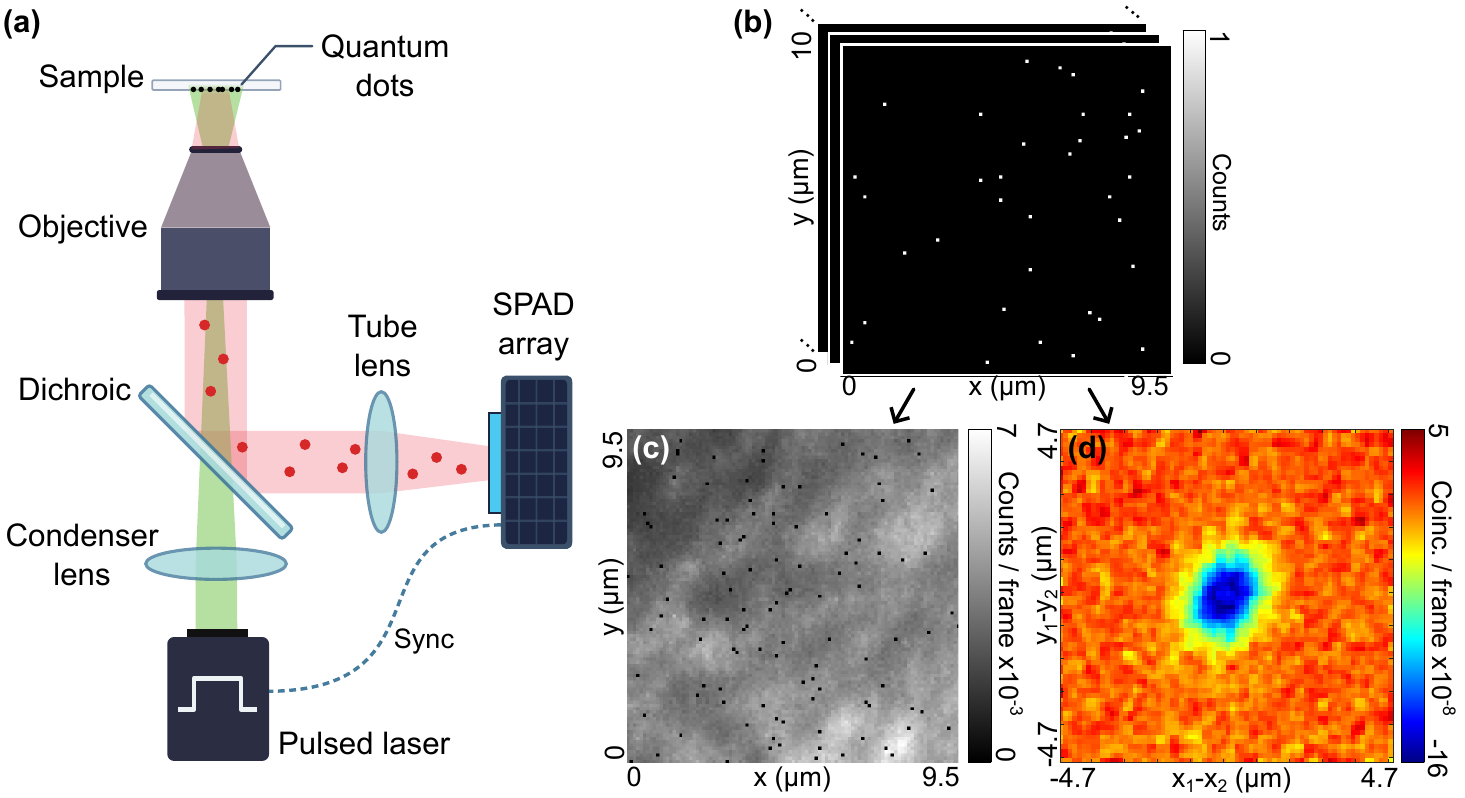}
\caption{\textbf{Concept of quantum-correlation-based aberration sensing}. \textbf{(a)} Schematic of the experimental setup based on a reflection-mode epi-fluorescence microscope. Quantum dots within the sample are excited by a pulsed laser ($\lambda = 520$ nm). The emitted fluorescence photons are collected by the objective and imaged onto a single-photon avalanche diode (SPAD) array synchronized with the excitation laser.
\textbf{(b)} Example of single frames acquired by the SPAD array with a $1 \mu s$ exposure time. For this experiment, optical aberrations were intentionally introduced into the imaging path.
\textbf{(c)} Conventional intensity image $I(\vec{r})$ obtained by summing $2\times10^5$ frames, where $\vec{r}=(x,y)$ denotes the transverse spatial coordinate in the detector plane. 
Individual quantum dots cannot be resolved due to the high density and the presence of severe optical aberrations. Hot pixels are set to zero for clarity.
\textbf{(d)} The spatial covariance image $C$ is calculated from the same set of frames. 
It represents the probability of detecting two photons in the same frame across any pair of pixels separated by a relative displacement vector of $\vec{r}_1-\vec{r}_2$, from which the probability of detecting them across the same pixel pair in distinct frames is subtracted.
The negative dip signifies the anti-bunching signature of single-photon emission from the quantum dots. The spatial shape of this dip directly reflects the aberrated point spread function (PSF) of the imaging system. }
\label{fig:setup}
\end{figure*}

Fortunately, most fluorophores used in bioimaging applications, such as fluorescent dyes, fluorescent proteins and quantum dots, exhibit quantum emission behavior: they emit photons one by one rather than in pairs, a phenomenon termed `antibunching'~\cite{Paul82_antibunching}. Essentially, fluorescent moieties can be treated as sources of ``missing photon pairs" (mathematically equivalent to correlated photon pairs generated by SPDC), whose absence can be inferred from photon correlation measurements using a Hanbury-Brown and Twiss (HBT) setup comprising a beamsplitter and two single photon detectors - or an equivalent architecture. Indeed, antibunching has been used in the past years for emitter counting~\cite{Weston2002,Grusmayer2019} and as a resource in super-resolution microscopy~\cite{Schwartz2013, Monticone2014, Israel2017, Tenne2019, Lubin:19}. The requirement to resolve single photons has historically confined these efforts to confocal imaging setups, using either individual avalanche photodiodes (APDs) or or small SPAD (single-photon avalanche diode) arrays as detectors.

The advent of large format SPAD arrays \cite{Bruschini2019} (with commercial systems up to 1 megapixel already available) has enabled the extension of these tools to wide-field imaging at reasonable integration times (seconds to minutes~\cite{Elmalem25} rather than hours if using EMCCDs or intensified CCDs~\cite{Schwartz2013}), thus making possible the use of antibunching as a resource for PSF characterization. 
Here we show that the PSF of a fluorescent microscope can indeed be characterized directly from the fluorescence image, without any additional restrictions or requirements, via the analysis of photon correlations across a wide field of view. This approach provides direct access even to a spatially varying PSF within the image. This characterization is achieved in a second or less, even with the limited capabilities of present-day SPAD arrays, and will likely be performed orders of magnitude faster - practically in real time - with next-generation devices.

\section{Theory} \label{sec:02_theory}

Let us consider a single fluorescent emitter located in the object plane at position $\vec{r_0}$. Upon excitation by a laser pulse, the emitted photons propagate through the imaging system and are recorded as clicks within a single SPAD camera frame (as shown in Fig. \ref{fig:setup}). We assume the low-flux regime, in which the mean number of photons arriving at any individual detector pixel is much smaller than unity. Under this assumption, the probability of registering a click at detector position $\vec r$ is given by
\begin{equation}
P_{\vec{r_0}}(\vec{r}) = \eta |h(\vec{r},\vec{r_0})|^2 O(\vec{r_0}),
\end{equation}
where $\eta$ is the overall detection efficiency (accounting for optical losses and the non-unity quantum efficiency of the detector), $h(\vec{r},\vec{r_0})$ is the point spread function (PSF), and $O(\vec{r_0})$ is the mean number of photons emitted by the source upon excitation by a pulse.

If the emitter can emit multiple photons within a single excitation cycle, a coincidence event at two distinct detector positions ($\vec r$) and ($\vec r'$) can be generated by two different photons emitted by the same source. In the same low-flux regime, the joint click probability is, to leading order,

\begin{equation}
P_{\vec{r_0}}(\vec{r}',\vec{r}) = \eta^2 |h(\vec{r},\vec{r_0})|^2 |h(\vec{r}',\vec{r_0})|^2 O(\vec{r_0})^2 (1 - \alpha(\vec{r_0})),
\end{equation}
where $\alpha(\vec{r_0})$ characterizes the purity of single-photon emission. In particular, ($\alpha=1$) for an ideal single-photon emitter, while ($\alpha=0$) for a Poissonian emitter (see supplemental document for details).

We now consider an ensemble of emitters distributed across the object plane, each emitting a mean photon number $O(\vec{r_0})$ upon excitation by a pulse.
In this case, the probability $P(\vec{r})$ of a detection event at the detector position $\vec{r}$ becomes

\begin{equation}
\label{eq1}
P(\vec{r}) = \int \eta |h(\vec{r},\vec{r_0})|^2 O(\vec{r_0}) d\vec{r_0}.
\end{equation}

Here, $O(\vec r_0)$ characterizes the spatial distribution and brightness of the emitters in the object. In particular, $O(\vec r_0)=0$ at positions where no emitter is present, while $O(\vec r_0)>0$ indicates the presence of an emitter.

A coincidence at positions $\vec r$ and $\vec r'$ can now arise in two ways: from photons emitted by two distinct emitters, or from two photons emitted by the same emitter. Assuming that different emitters emit independently, the joint coincidence probability is

\begin{equation}
\label{eq2}
\begin{split}
    P(\vec{r}',\vec{r}) =& \iint \eta^2 |h(\vec{r},\vec{r}_0)|^2 |h(\vec{r}',\vec{r}_1)|^2 \\
    & O(\vec{r}_0) O(\vec{r}_1) \left[1 - \alpha(\vec{r}_0) \delta(\vec{r}_0-\vec{r}_1)\right] d\vec{r}_0 d\vec{r}_1,
\end{split}
\end{equation}
where $\delta$ is the Dirac delta function. The first term in the square brackets describes coincidences generated by two independent emitters. The second term accounts for the purity of the individual emitters.

In our experiment, we extract the photon correlation information by computing the spatial covariance image $C(\vec{r}) = \langle S \star S \rangle (\vec{r}) - \left[ \langle S \rangle \star \langle S \rangle (\vec{r}) \right]$, where $S(\vec{r})$ is the binary random variable ($0$ or $1$) representing a detection event recorded at position $\vec{r}$ of a single frame (see Fig. \ref{fig:setup}(b)). The angular brackets $\langle \dots \rangle$ are the ensemble average over many excitations. The correlation operation $\star$ is defined as $S \star S (\vec{r}) =  \Sigma _p S(p) S(p-\vec{r})$.
Using Equations~\eqref{eq1} and~\eqref{eq2} and assuming that the PSF is shift-invariant over the region of interest (i.e. $h(\vec{r}',\vec{r}) = h(\vec{r}'-\vec{r})$), the covariance simplifies to
\begin{equation}
\label{eq:psf_corr}
C(\vec{r}) = - K [|h|^2 \star |h|^2 ] (\vec{r}),
\end{equation}
where $K = \eta^2 \int O(\vec{r})^2 \alpha(\vec{r}) d\vec{r}$ is a constant. 

\Cref{eq:psf_corr} is our central result: the two-point photon correlation function, detected here via a measurement of the spatial covariance $C(\vec{r})$ at the single-photon level, is proportional to the autocorrelation of the incoherent PSF, scaled by a constant prefactor determined jointly by the object's structure, emitter statistics, and the overall detection efficiency. 
Notably, $C$ corresponds exactly to the projection of the connected correlation function $G^{(2)}(\vec{r}_1,\vec{r}_2,\tau=0) - G^{(1)}(\vec{r}_1,\tau=0)G^{(1)}(\vec{r}_2,\tau=0)$ onto the minus-coordinate axis $\vec{r}_1-\vec{r}_2$, where $G^{(2)}$ and $G^{(1)}$ are the second- and first-order intensity correlation functions at zero delay~\cite{defienne2018general} (see supplemental document for detailed derivations).

\section{Experiment} \label{sec:04_exp}

The experimental setup is described in \Cref{fig:setup}(a). Our scheme requires minimal changes to a conventional fluorescence microscope: replace the detector with a SPAD array (operated in single-photon binary sensing mode) and excite the sample using one short illumination pulse per frame. 
For quantum emitters, this results in single-photon per emitter per pulse, allowing to acquire the quantum correlations of the emitted light. \Cref{fig:setup}(b) shows an individual frame in the experiment, with the white dots representing the detection of individual photons. 

The average of many such frames yields the intensity image $I$ (\Cref{fig:setup}(c)), which is also accessible via classical measurements. 
Furthermore, the spatial covariance image $C$ can be calculated from this same set of frames. 
To do so, we first compute the spatial autocorrelation of each individual frame and average these results over the entire set. 
From this averaged term, we then subtract the spatial autocorrelation of the overall intensity image. 

In \Cref{fig:setup}(c), the negative values in the central region confirm the single-photon (antibunched) nature of our individual emitters. Furthermore, as described by \Cref{eq:psf_corr}, measuring the spatial covariance effectively distills the PSF information by removing any object dependence. In this example, the spatial profile of the negative region is visibly distorted and broadened, directly confirming the presence of aberrations within the imaging system.

Detailed description of the experimental setup, sample preparation and data processing is provided in the supplemental document. The code for the spatial $\Gt$ calculation is available at [will be uploaded to GitHub upon publication] and the temporal $\gt$ calculation code is available at \cite{g2Code}.

\section{Results} \label{sec:05_res}

As a first demonstration of the proposed method's capabilities, we evaluated defocus aberrations by performing a focal-stack measurement. 
The sample was translated axially around the microscope's focal plane through several discrete positions, with $N=200,000$ binary frames acquired at each step. 
A subset of these results is presented in \Cref{fig:results_cyl}(a-j) (see supplemental document for the complete dataset and Visualization 1 for video displaying the results). 
A clear correlation between the intensity image sharpness and the PSF width can be observed. 

\begin{figure*}[t]
\centering
\includegraphics[width=1\linewidth]{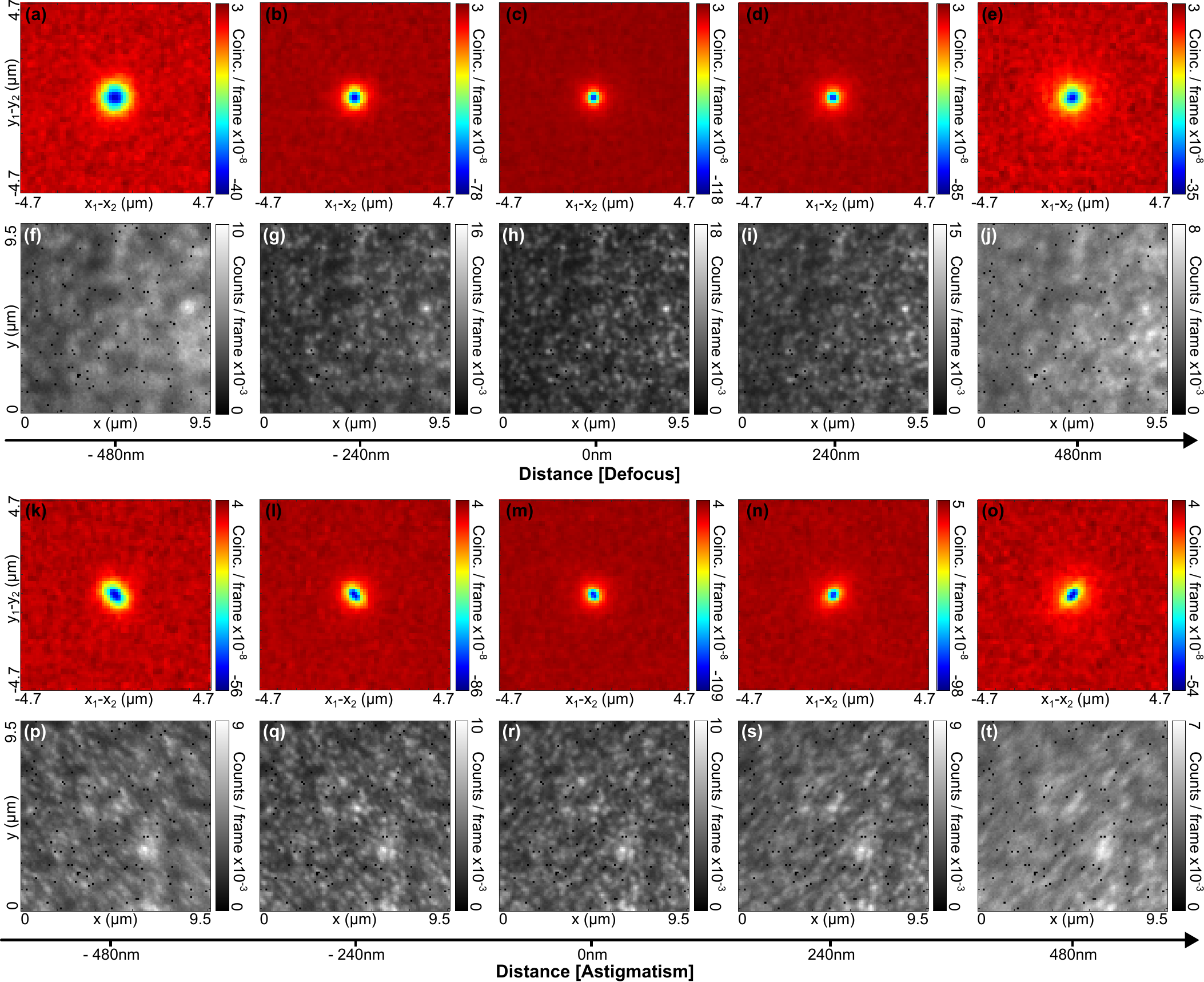}
\caption{\label{fig:results_cyl} \textbf{Experimental results for defocus and astigmatism aberrations.} To evaluate defocus, the sample is translated axially across the focal plane. 
Panels \textbf{(a-e)} display the resulting spatial covariance images ($C$), and \textbf{(f-j)} show the corresponding intensity images ($I$). To induce astigmatism, a cylindrical lens is introduced into the imaging path, rendering the PSF elliptical. As the sample is translated across the focal plane, the orientation of the elliptical PSF rotates by $90$ degrees, providing a distinct signature of the aberration. Panels \textbf{(k-o)} present the corresponding spatial covariance images, and \textbf{(p-t)} display the resulting intensity images (see Visualizations 1-2 for the full results in a video format).}
\end{figure*}

Then, a cylindrical lens with $f=500 mm$ was added to the optical path, roughly $d=50 mm$ before the sensor, to evaluate the method under astigmatic aberrations. 
This addition, along with similar focus-sweep and acquisition settings, introduced a strong astigmatism aberration around the nominal focus point. 
A subset of these results are presented in \Cref{fig:results_cyl}(k-t) (full results in supplemental document and Visualization 2). 
Astigmatism is clearly observed in the PSF estimation, as well as in the corresponding intensity images.

\begin{figure*}[t]
\centering
\includegraphics[width=1\linewidth]{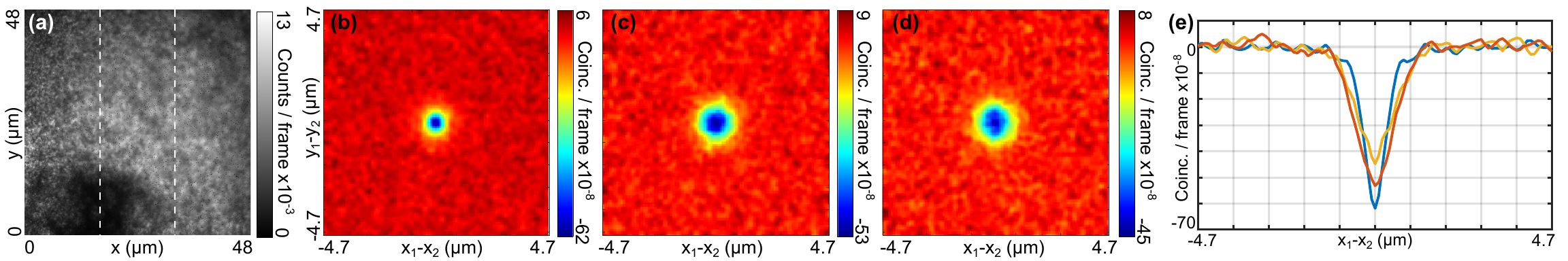}
\caption{\label{fig:results_shift_var} \textbf{Experimental results for a shift-variant aberration.} The sample was tilted along a transverse axis such that different parts of the sample experienced varying defocus, creating a shift-variant PSF. \textbf{(a)} Average intensity measurement within the full field of view (FOV). The left portion of the image appears more focused than the right portion. To reveal the shift-varying PSF, we calculate the spatial covariance $C(\vec{r})$ independently for three segments of the FOV, denoted by dashed lines in (a). \textbf{(b-d)} Spatial covariance images for the left, center, and right segments, clearly showing the increasing defocus of the PSF. \textbf{(e)} 1D cross-sections at $y_1-y_2=0$ of the left (blue), center (red), and right (yellow) covariance images, illustrating both the broadening and the reduction in magnitude of the central dip (see Visualizations 3-4 for the full results in a video format). 
}
\end{figure*}

Since aberrations are spatially variant in many practical cases, a scenario with spatially varying defocus was generated by tilting the sample. 
A similar focus-stack was acquired, and the PSF estimation was carried on different regions of interest in the image (three vertical stripes), as shown in \Cref{fig:results_shift_var}(a).
Covariances images for three stripes are presented in \Cref{fig:results_shift_var}(b-d) (see supplemental document and Visualization 3 for the complete dataset). 
Although the estimation relies on only one-third of the data, the method is sufficiently robust to yield a high-quality PSF estimation without increasing the number of frames. 
The effect of the sample tilt across the focus sweep is clearly visible in the 1D cross-sections through the center of the covariance matrix presented in \Cref{fig:results_shift_var}(e). 
Additional results analyzing the spatio-temporal SNR tradeoff when splitting the same dataset to five vertical stripes are presented in the supplemental document and Visualization 4.

\begin{figure}[t]
\centering
\includegraphics[width=1 \columnwidth]{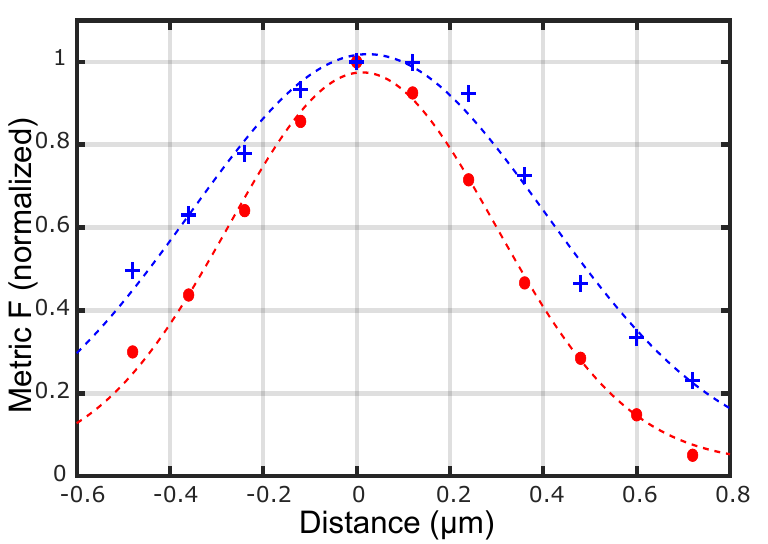}
\caption{\label{fig:results_curve} \textbf{Dependence of the metric $F$ on aberration strength.} Values of the metric $F$ (calculated using the temporal $\gt$ option), are plotted as a function of the sample's axial displacement from the focal plane for the defocus (red circles) and defocus with astigmatism (blue crosses) cases. 
For visualization purposes, the values of $F$ have been normalized to their maximum.
Larger axial displacements (i.e. moving the sample further from the focal plane) correspond to stronger induced aberrations. 
The blue and red dashed lines represent Gaussian fits.}
\end{figure}

Extracting the PSF via our technique allows us to estimate and compensate for optical aberrations using adaptive optics (AO). 
To this end, we define a feedback metric $F = -  C(\vec{r} = \vec{0})$. The metric reaches its maximum when aberrations are absent (i.e. when the object is perfectly in focus), demonstrating its potential for AO feedback. 
A key advantage of this metric is its complete independence from the object's morphology. Practically, this metric can also be expressed via the normalized second-order correlation function: $F = \int I(\vec{r})^2 [1 - g^{(2)}(\vec{r},\vec{r},\tau=0)] d\vec{r}$. Thus, retrieving it does not require to compute the full correlation function or spatial covariance. It simply requires detecting photon coincidences at each pixel and averaging them across the array - a localized scheme already used in super-resolution approaches~\cite{Elmalem25}. 
In practice, because most sensors cannot resolve same-pixel coincidences, this quantity is estimated by averaging coincidences between immediate neighboring pixels (see \cite{Elmalem25}).
The curves in \Cref{fig:results_curve} show the behavior of this metric for the defocus and astigmatism aberrations measured in \Cref{fig:results_cyl}.

\section{Discussion and Conclusion} \label{sec:06_disc}

The presented method and results hold potential for enabling closed-loop auto-focus and adaptive optics, as well as post-processing PSF-based deconvolution. 
All are outcomes of the quantum nature of the fluorescent emitters and the single-photon sensitivity of the SPAD array, eliminating the requirement for a wavefront sensor, phase-detection pixels, or guide-star like objects in the sample. In order to assess the practicality of the method, several aspects should be discussed.

Regarding acquisition speed, our method already demonstrates highly efficient data utilization. 
While our standard measurements use $N=200,000$ frames (i.e. $\sim2$ seconds of acquisition), robust PSF estimations are readily achievable with a third or a fifth of this dataset (\Cref{fig:results_shift_var} and supplemental document). 
By further tuning experimental tradeoffs (excitation power, quantum dots saturation and bleaching, sensor saturation), usable signals can even be extracted from as few as $N=10,000$ frames, corresponding to $\sim0.1$ seconds (see supplemental document).
Acquisition times could be further reduced by bypassing the $100 kHz$ read-out limit of the SPAD512 camera, as intrinsic pixel dead times support frame rates $>10 MHz$. 
Furthermore, shifting photon coincidence calculations to on-chip architectures \cite{Gorman24} would eliminate read-out latency and provide instantaneous access to the feedback metric $F$. 
This would enable real-time aberration sensing at rates exceeding $1 kHz$, unlocking fast closed-loop adaptive optics capable of correcting the dynamic aberrations typically encountered in vivo.

Regarding the spatial covariance measurement, as formulated in \cref{sec:02_theory}, it yields the auto-correlation of the incoherent PSF. 
While this does not provide a direct wavefront measurement, it preserves a fundamental property: the correlation spot is narrowest when aberrations are completely compensated. 
Consequently, alternative metrics beyond $F$ could exploit additional features of the covariance image, such as its width, to drive more sophisticated adaptive optics strategies \cite{Katz:14}. 
Furthermore, the PSF can be reconstructed from its auto-correlation via inverse problem solvers \cite{fienup1982phase}, and subsequently applied to deconvolve the aberrated image in post-processing.

Regarding fluorescent agent compatibility, our method fundamentally relies on a single-photon per-emitter per-frame regime. 
While we predominantly used bright and stable quantum dots, standard bioimaging fluorophores like organic dyes and fluorescent proteins are equally compatible, despite their faster photobleaching rates. 
Consequently, the main constraint of our method is not the nature of emitter, but its density. 
Because binary sensors saturate at a single photon per pixel per frame, dense samples - particularly those employing secondary labeling  - can easily induce detector saturation. 
This problem is currently mitigated by increasing optical magnification to distribute photons across more pixels, albeit at the expense of a smaller field of view. 
Fortunately, as the adoption of on-chip processing alleviates current data-output limits, SPAD arrays will rapidly scale in size and naturally resolve this density limit. 
Ultimately, the wavefront correction offered by our approach, combined with synergistic SPAD capabilities like per-pixel $g^{(2)}$ evaluation and super-resolution \cite{Elmalem25}, provides a highly compelling tradeoff for future imaging applications.

Overall, this work establishes the intrinsic single-photon emission of conventional fluorescent emitters as a practical resource for aberration sensing in fluorescence microscopy. With larger SPAD arrays incorporating on-chip processing, this approach could provide a simple solution for real-time aberration sensing in conventional fluorescence microscopes. Beyond the specific implementation demonstrated here, these results highlight the potential of photon correlation sensing to enable new modalities that can be incorporated into conventional imaging systems with minimal hardware modifications.

\begin{backmatter}
\bmsection{Funding}
[PlaceHolder for auto-genreated funding section]

\bmsection{Acknowledgment}
D.O. acknowledges funding from the Israel Science Foundation (grant No. 1249/25). D.O. and H.D. acknowledge support by the International Research Network OMNI-TOOLS. H.D. acknowledges funding from the ERC Starting Grant (No. SQIMIC-101039375). D.O. is the incumbent of the Harry Weinrebe professorial chair of laser physics.

\bmsection{Disclosures}
[PlaceHolder TBD]

\bmsection{Data availability} Data underlying the results presented in this paper are not publicly available at this time but may be obtained from the authors upon reasonable request.

\bmsection{Supplemental document}
See Supplement 1 for full theoretical derivation and additional results, and Visualizations 1-4 for videos displaying the results.

\end{backmatter}

\bibliography{bib}

@article{defienne2018general,
  title={General model of photon-pair detection with an image sensor},
  author={Defienne, Hugo and Reichert, Matthew and Fleischer, Jason W},
  journal={Physical review letters},
  volume={120},
  number={20},
  pages={203604},
  year={2018},
  publisher={APS}
}

@article{fienup1982phase,
  title={Phase retrieval algorithms: a comparison},
  author={Fienup, James R},
  journal={Applied optics},
  volume={21},
  number={15},
  pages={2758--2769},
  year={1982},
  publisher={Optical Society of America}
}

@ARTICLE{Israel2017,
  title     = "Quantum correlation enhanced super-resolution localization
               microscopy enabled by a fibre bundle camera",
  author    = "Israel, Yonatan and Tenne, Ron and Oron, Dan and Silberberg,
               Yaron",
  journal   = "Nat. Commun.",
  publisher = "Springer Science and Business Media LLC",
  volume    =  8,
  number    =  1,
  pages     = "14786",
  month     =  mar,
  year      =  2017,
  copyright = "https://creativecommons.org/licenses/by/4.0"
}

@article{Schwartz2013,
author = {Schwartz, Osip and Levitt, Jonathan M. and Tenne, Ron and Itzhakov, Stella and Deutsch, Zvicka and Oron, Dan},
title = {Superresolution Microscopy with Quantum Emitters},
journal = {Nano Letters},
volume = {13},
number = {12},
pages = {5832-5836},
year = {2013},
doi = {10.1021/nl402552m},
    note ={PMID: 24195698},

URL = { 
    
        https://doi.org/10.1021/nl402552m
    
    

},
eprint = { 
    
        https://doi.org/10.1021/nl402552m
    
    

}
}

@article{Elmalem25,
author = {Shay Elmalem and Gur Lubin and Michael Wayne and Claudio Bruschini and Edoardo Charbon and Dan Oron},
journal = {Optica},
number = {4},
pages = {451--458},
publisher = {Optica Publishing Group},
title = {Massively multiplexed wide-field photon correlation sensing},
volume = {12},
month = {Apr},
year = {2025},
url = {https://opg.optica.org/optica/abstract.cfm?URI=optica-12-4-451},
doi = {10.1364/OPTICA.550498},
}

@ARTICLE{Defienne2024,
  title     = "Advances in quantum imaging",
  author    = "Defienne, Hugo and Bowen, Warwick P and Chekhova, Maria and
               Lemos, Gabriela Barreto and Oron, Dan and Ramelow, Sven and
               Treps, Nicolas and Faccio, Daniele",
  journal   = "Nat. Photonics",
  publisher = "Springer Science and Business Media LLC",
  volume    =  18,
  number    =  10,
  pages     = "1024--1036",
  month     =  oct,
  year      =  2024,
  copyright = "https://www.springernature.com/gp/researchers/text-and-data-mining"
}

@article{
Cameron2024,
author = {Patrick Cameron  and Baptiste Courme  and Chloé Vernière  and Raj Pandya  and Daniele Faccio  and Hugo Defienne },
title = {Adaptive optical imaging with entangled photons},
journal = {Science},
volume = {383},
number = {6687},
pages = {1142-1148},
year = {2024},
doi = {10.1126/science.adk7825},
URL = {https://www.science.org/doi/abs/10.1126/science.adk7825},
eprint = {https://www.science.org/doi/pdf/10.1126/science.adk7825}}

@ARTICLE{Monticone2014,
  title     = "Beating the Abbe diffraction limit in confocal microscopy via
               nonclassical photon statistics",
  author    = "Gatto Monticone, D and Katamadze, K and Traina, P and Moreva, E
               and Forneris, J and Ruo-Berchera, I and Olivero, P and
               Degiovanni, I P and Brida, G and Genovese, M",
  journal   = "Phys. Rev. Lett.",
  publisher = "American Physical Society (APS)",
  volume    =  113,
  number    =  14,
  pages     = "143602",
  month     =  oct,
  year      =  2014,
  copyright = "http://link.aps.org/licenses/aps-default-license"
}

@ARTICLE{Tenne2019,
  title     = "Super-resolution enhancement by quantum image scanning
               microscopy",
  author    = "Tenne, Ron and Rossman, Uri and Rephael, Batel and Israel,
               Yonatan and Krupinski-Ptaszek, Alexander and Lapkiewicz, Radek
               and Silberberg, Yaron and Oron, Dan",
  journal   = "Nat. Photonics",
  publisher = "Springer Science and Business Media LLC",
  volume    =  13,
  number    =  2,
  pages     = "116--122",
  month     =  feb,
  year      =  2019
}

@ARTICLE{Grusmayer2019,
  title     = "Photons in - numbers out: perspectives in quantitative
               fluorescence microscopy for in situ protein counting",
  author    = "Gru$\beta$mayer, K S and Yserentant, K and Herten, D-P",
  journal   = "Methods Appl. Fluoresc.",
  publisher = "IOP Publishing",
  volume    =  7,
  number    =  1,
  pages     = "012003",
  month     =  jan,
  year      =  2019,
  copyright = "https://publishingsupport.iopscience.iop.org/iop-standard/v1"
}

@article{Weston2002,
author = {Weston, Kenneth D. and Dyck, Martina and Tinnefeld, Philip and M{\"u}ller, Christian and Herten, Dirk P. and Sauer, Markus},
title = {Measuring the Number of Independent Emitters in Single-Molecule Fluorescence Images and Trajectories Using Coincident Photons},
journal = {Analytical Chemistry},
volume = {74},
number = {20},
pages = {5342-5349},
year = {2002},
doi = {10.1021/ac025730z},
    note ={PMID: 12403591},

URL = { 
    
        https://doi.org/10.1021/ac025730z
    
    

},
eprint = { 
    
        https://doi.org/10.1021/ac025730z
    
    

}

}

@ARTICLE{Bruschini2019,
  title     = "Single-photon avalanche diode imagers in biophotonics: review
               and outlook",
  author    = "Bruschini, Claudio and Homulle, Harald and Antolovic, Ivan
               Michel and Burri, Samuel and Charbon, Edoardo",
  journal   = "Light Sci. Appl.",
  publisher = "Springer Science and Business Media LLC",
  volume    =  8,
  number    =  1,
  pages     = "87",
  month     =  sep,
  year      =  2019,
  copyright = "https://creativecommons.org/licenses/by/4.0"
}

@article{Johnson2024,
author = {Courtney Johnson and Min Guo and Magdalena C. Schneider and Yijun Su and Satya Khuon and Nikolaj Reiser and Yicong Wu and Patrick La Riviere and Hari Shroff},
journal = {Optica},
number = {6},
pages = {806--820},
publisher = {Optica Publishing Group},
title = {Phase-diversity-based wavefront sensing for fluorescence microscopy},
volume = {11},
month = {Jun},
year = {2024},
url = {https://opg.optica.org/optica/abstract.cfm?URI=optica-11-6-806},
doi = {10.1364/OPTICA.518559},
}

@ARTICLE{Ji2010,
  title     = "Adaptive optics via pupil segmentation for high-resolution
               imaging in biological tissues",
  author    = "Ji, Na and Milkie, Daniel E and Betzig, Eric",
  journal   = "Nat. Methods",
  publisher = "Springer Science and Business Media LLC",
  volume    =  7,
  number    =  2,
  pages     = "141--147",
  month     =  feb,
  year      =  2010
}

@article{Azucena2010,
author = {Oscar Azucena and Justin Crest and Jian Cao and William Sullivan and Peter Kner and Donald Gavel and Daren Dillon and Scot Olivier and Joel Kubby},
journal = {Opt. Express},
number = {16},
pages = {17521--17532},
publisher = {Optica Publishing Group},
title = {Wavefront aberration measurements and corrections through thick tissue using fluorescent microsphere reference beacons},
volume = {18},
month = {Aug},
year = {2010},
url = {https://opg.optica.org/oe/abstract.cfm?URI=oe-18-16-17521},
doi = {10.1364/OE.18.017521},
}

@article{Tao2012,
author = {Xiaodong Tao and Justin Crest and Shaila Kotadia and Oscar Azucena and Diana C. Chen and William Sullivan and Joel Kubby},
journal = {Opt. Express},
number = {14},
pages = {15969--15982},
publisher = {Optica Publishing Group},
title = {Live imaging using adaptive optics with fluorescent protein guide-stars},
volume = {20},
month = {Jul},
year = {2012},
url = {https://opg.optica.org/oe/abstract.cfm?URI=oe-20-14-15969},
doi = {10.1364/OE.20.015969},
}

@ARTICLE{Hampson2021,
  title     = "Adaptive optics for high-resolution imaging",
  author    = "Hampson, Karen M and Turcotte, Rapha{\"e}l and Miller, Donald T
               and Kurokawa, Kazuhiro and Males, Jared R and Ji, Na and Booth,
               Martin J",
  journal   = "Nat. Rev. Methods Primers",
  publisher = "Springer Science and Business Media LLC",
  volume    =  1,
  number    =  1,
  pages     = "68",
  month     =  oct,
  year      =  2021,
  copyright = "https://www.springernature.com/gp/researchers/text-and-data-mining"
}

@article{Gorman24,
author = {Alistair Gorman and Neil Finlayson and Ahmet T. Erdogan and Lars Fisher and Yining Wang and Francescopaolo Mattioli Della Rocca and Hanning Mai and Edbert J. Sie and Francesco Marsili and Robert K. Henderson},
journal = {Biomed. Opt. Express},
number = {11},
pages = {6499--6515},
publisher = {Optica Publishing Group},
title = {ATLAS: a large array, on-chip compute SPAD camera for multispeckle diffuse correlation spectroscopy},
volume = {15},
month = {Nov},
year = {2024},
url = {https://opg.optica.org/boe/abstract.cfm?URI=boe-15-11-6499},
doi = {10.1364/BOE.531416},
}

@misc{SA_560_QDs,
  author = {Sigma-Aldrich},
  title = {CdSe/ZnS core-shell type quantum dots},
  year = 2026,
  url = {https://www.sigmaaldrich.com/IL/en/product/aldrich/919071?srsltid=AfmBOoohbQL6zxgrNrT8SLviWNX8RpPXvh7hXjqYMknxVwdr-dC4kXfS},
  urldate = {2026-06-28}
}

@misc{SPAD_512,
  author = {Pi-imaging},
  title = {SPAD 512},
  year = 2026,
  url = {https://piimaging.com/spad-512/},
  urldate = {2026-06-28}
}

@misc{g2Code,
  author = {Shay Elmalem},
  title = {{WF-g2} Wide-field photon correlation calculation},
  howpublished = {\url{https://github.com/ShayElmalem/WF-g2}},
  year = 2025,
  note = {Accessed: 2026-06-29}
}

@article{Paul82_antibunching,
  title = {Photon antibunching},
  author = {Paul, H.},
  journal = {Rev. Mod. Phys.},
  volume = {54},
  issue = {4},
  pages = {1061--1102},
  numpages = {0},
  year = {1982},
  month = {Oct},
  publisher = {American Physical Society},
  doi = {10.1103/RevModPhys.54.1061},
  url = {https://link.aps.org/doi/10.1103/RevModPhys.54.1061}
}

@article{Lubin:19,
author = {Gur Lubin and Ron Tenne and Ivan Michel Antolovic and Edoardo Charbon and Claudio Bruschini and Dan Oron},
journal = {Opt. Express},
number = {23},
pages = {32863--32882},
publisher = {Optica Publishing Group},
title = {Quantum correlation measurement with single photon avalanche diode arrays},
volume = {27},
month = {Nov},
year = {2019},
url = {https://opg.optica.org/oe/abstract.cfm?URI=oe-27-23-32863},
doi = {10.1364/OE.27.032863},
}

@article{Katz:14,
author = {Ori Katz and Eran Small and Yefeng Guan and Yaron Silberberg},
journal = {Optica},
number = {3},
pages = {170--174},
publisher = {Optica Publishing Group},
title = {Noninvasive nonlinear focusing and imaging through strongly scattering turbid layers},
volume = {1},
month = {Sep},
year = {2014},
url = {https://opg.optica.org/optica/abstract.cfm?URI=optica-1-3-170},
doi = {10.1364/OPTICA.1.000170},
}

\end{document}


\maketitle

\section{Extended theoretical model}
\label{sec:supp-theory}

Following Section II of the main paper, we present the detailed derivation of the theoretical model.
\subsection{Derivation of the joint probability for a single emitter}
We derive the joint click probability $P_{\vec{r_0}}(\vec{r}',\vec{r})$ for a single emitter located at position $\vec{r_0}$. Let ($m$) denote the number of photons emitted by the source following a single excitation pulse. In the low-flux regime, where the probability of multiple photons arriving at the same detector pixel is negligible, the probability that one photon is detected at position ($\vec r$) and a second photon is detected at position ($\vec r'$), conditioned on the emission of ($m$) photons, is

\begin{equation}
\label{eqn_sm_cond}
P_{\vec r_0}(\vec r',\vec r|m)
\simeq m(m-1) \eta^2 |h(\vec r,\vec r_0)|^2 |h(\vec r',\vec r_0)|^2.
\end{equation}

The factor ($m(m-1)$) counts the number of ordered pairs of distinct photons that can give rise to the two detections. The factor ($\eta^2$) accounts for the detection efficiency of the two photons, while the two PSF factors describe the probability that the photons are detected at the respective detector positions.

We then average Eq. (\ref{eqn_sm_cond}) over the photon-number distribution ($P_{\vec r_0}(m)$) of the emitter. Defining the mean number of photons emitted per pulse $ \langle m\rangle_{\vec r_0} = O(\vec r_0)$, and the correlations $\left\langle m(m-1)\right\rangle = O(\vec r_0)^2g^{(2)}(\vec r_0,\tau = 0)$, we get

\begin{equation}
\label{eqn_sm_joint}
P_{\vec r_0}(\vec r',\vec r) = \eta^2 |h(\vec r,\vec r_0)|^2
|h(\vec r',\vec r_0)|^2 O(\vec r_0)^2 g^{(2)}(\vec r_0,\tau = 0).
\end{equation}

\noindent where $g^{(2)}(\vec r_0,\tau = 0)$ is the normalized second order correlation function at time delay $\tau = 0$. In the main text, we defined the parameter $\alpha(\vec r_0) = 1- g^{(2)}(\vec r_0,\tau = 0)$ which characterizes the purity of single-photon emission.

\subsection{Covariance as minus-coordinate projection of $G^{(2)}$}

In the derivation of the covariance $C(\vec{r})$, we mentioned that it is exactly equal to the minus-coordinate projection of the connected correlations $G^{(2)}(\vec{r}_1,\vec{r}_2,\tau=0) - G^{(1)}(\vec{r}_1,\tau=0)G^{(1)}(\vec{r}_2,\tau=0)$. We show its derivation here.

The second-order spatial correlation function (at $\tau = 0$) of the optical field for $\vec r_1 \neq \vec r_2$, is described in terms of the photon number operator $\hat n(\vec r)$, such that, $G^{(2)}(\vec{r}_1,\vec{r}_2,\tau=0) = \langle \hat n(\vec r_1)\hat n(\vec r_2) \rangle$. In the low-flux regime, where the probability of two or more photons arriving at the same detector pixel is negligible, we define a random variable $S(\vec{r})$ to model the photon detection by our camera. It is a binary random variable ($0$ or $1$) representing a detection event recorded at position $\vec{r}$ of the detector. Thus, the correlation function can be written as  $G^{(2)}(\vec{r}_1,\vec{r}_2,\tau=0) = \langle S(\vec{r}_1)S(\vec{r}_2) \rangle$. Similarly, the first-order correlation function (intensity) becomes $G^{(1)}(\vec{r},\tau=0) = \langle S(\vec{r})\rangle$. 

Projecting the second order correlation function along the minus coordinate $(\vec{r_-} = \vec{r}_2 - \vec{r}_1)$, we get $\Sigma_{\vec{r}_+} \langle S(\vec{r}_+ - \vec{r}_-/2)S(\vec{r}_+ + \vec{r}_-/2) \rangle$, where $\vec{r}_+ = (\vec{r}_2 + \vec{r}_1)/2$ and $\vec{r}_- = (\vec{r}_2 - \vec{r}_1)$. This term is exactly equal to $\langle S \star S \rangle (\vec{r}) = \langle \Sigma _p S(p) S(p-\vec{r_-}) \rangle $ of the main text. Similarly, the minus-coordinate projection of the product of intensities becomes $\Sigma_{\vec{r}_+} G^{(1)}(\vec{r}_+ - \vec{r}_-/2,\tau=0)G^{(1)}(\vec{r}_+ + \vec{r}_-/2,\tau=0) =  \langle S \rangle \star \langle S \rangle (\vec{r_-})$, yielding the covariance function described in the main text.

\section{Setup and experimental process details}
\label{sec:supp-setup}

\subsection{Experimental setup}
The experiments were performed using a conventional widefield epi-fluorescent microscope setup, with a sub-\SI{}{\pico\second} pulsed laser as an illumination source, and a SPAD array detector as the camera. A standard inverted microscope (Zeiss Axiovert 35) serves as a microscope body. A \SI{300}{\femto\second} pulsed laser (Spirit One, SpOne-8-F2P-SHG by Spectra Physics) with a central wavelength of \SI{520}{\nano\meter} is used as an illumination source. The laser light is focused in the back focal plane of the objective (Zeiss 440285, 100x, $NA=1.3$) to generate widefield illumination in the sample plane. The light is collected by the same objective, and the emission is filtered from the excitation light using a dichroic mirror (Semrock FF552-Di02) and an additional filter (Semrock BLP01-532R-25). The light is then focused on the SPAD 512 \cite{SPAD_512} located in the lower camera port of the microscope. An additional magnification in this port yields a final magnification of $M=175$. The sample is located on a piezo stage (Mad City Labs MCL-MANNZ) enabling focus-sweep acquisition.

\subsection{Sample preparation}
The samples used in this work are based on a spin-coated diluted solution of CdSe/ZnS core-shell type quantum-dots (QDs) with emission centered around $\lambda_{em}=560 nm$ \cite{SA_560_QDs}. The stock QD solution was diluted at a ratio of 1:500 in a 3\% (w/v) PMMA in toluene solution. The diluted solution is then spin-coated on $\#0$ glass coverslip. 

\subsection{Data acquisition and analysis}
The SPAD 512 is operated in its fast 1-bit imaging mode, enabling acquisition in short bursts with frame rate of up to \SI{100}{\kilo\hertz}. The laser is operated in \SI{70}{\kilo\hertz}, and its average output power is \SI{2.6}{\milli\watt} in all the experiments presented in the paper and the SI, besides the experiment presented in \cref{fig:sm_results_power5x} in which the average power is \SI{13}{\milli\watt}. To synchronize the SPAD frames to the laser pulses, the laser output is tipped-off and sensed using a photo-diode (New Focus 2031), and its output is being used as a synchronization signal for the SPAD frames. A \SI{1}{\micro\second} single gate is opened after the laser pulse, to allow for the emission's detection and to decrease the acquired dark-counts (note that while gates of down to \SI{5}{\nano\second} are possible, various uncertainties in the synchronization mechanism led us to increase the gate width to \SI{1}{\micro\second}). The data is acquired in bursts of $N=10k$ frames, which are transferred to a computer and processed offline. 

Before performing the covariance $C(\vec{r})$ calculation, a pre-processing step is applied to remove hot pixels using a hot-pixel map estimated from a dark measurement. The acquired binary frames are then processed to estimate the PSF autocorrelation through the spatial covariance $C(\vec{r})$, as described in Section 2 of the main paper. The first term of the covariance, $\langle S \star S \rangle(\vec{r})$, is obtained by summing the autocorrelations of individual frames. The second term, $\langle S \rangle \star \langle S \rangle(\vec{r})$, is estimated by summing the cross-correlations between consecutive frames \cite{defienne2018general}. The correlation was computed using MATLAB’s \textit{xcorr2()} function, and the resulting matrix was normalized by the autocorrelation of a matrix of ones to remove zero-padding artifacts.

To compensate for crosstalk, independent crosstalk probability estimation (using the same process described in \cite{Elmalem25}) was performed. Very low crosstalk probability of $p_{CT}=2.6\cdot10^{-4}$ was observed for the four edge-sharing neighbors, and negligible crosstalk for any other neighbor. Crosstalk pairs were estimated statistically by $CT=\sum p_{CT}*I$ (where $I$ is the intensity counts) and reduced from the relevant pixels (note that in the current SNR and sampling regime, exact crosstalk estimation can be avoided and such correction can be performed with simple interpolation which leads to similar results). As the central pixel includes the zero-shift auto-correlation, its value is replaced with four-neighbors interpolation. The covariance matrix is subsequently smoothed using a Gaussian filter with a standard deviation of one pixel. The central $101 \times 101$ pixel region is then cropped to generate the images presented in this paper. For the intensity images, hot pixels in the averaged intensity frames are set to zero for improved visualization, and a central $101 \times 101$ pixel region is cropped from the full $512 \times 512$ field of view (except for Fig.~3).

In addition to the spatial $\Gt$, temporal $\gt$ for the four edge-sharing neighbors is calculated as well (with crosstalk and dark-counts compensation, as described in \cite{Elmalem25}), and then used for the focus measure calculation. 
The code for the spatial $\Gt$ calculation is available at [will be uploaded to GitHub upon publication] and the temporal $\gt$ calculation code is available at \cite{g2Code}.

\section{Additional Experimental Results}
\label{sec:supp-results}

\begin{figure*}[t]
\centering
\includegraphics[width=1\linewidth]{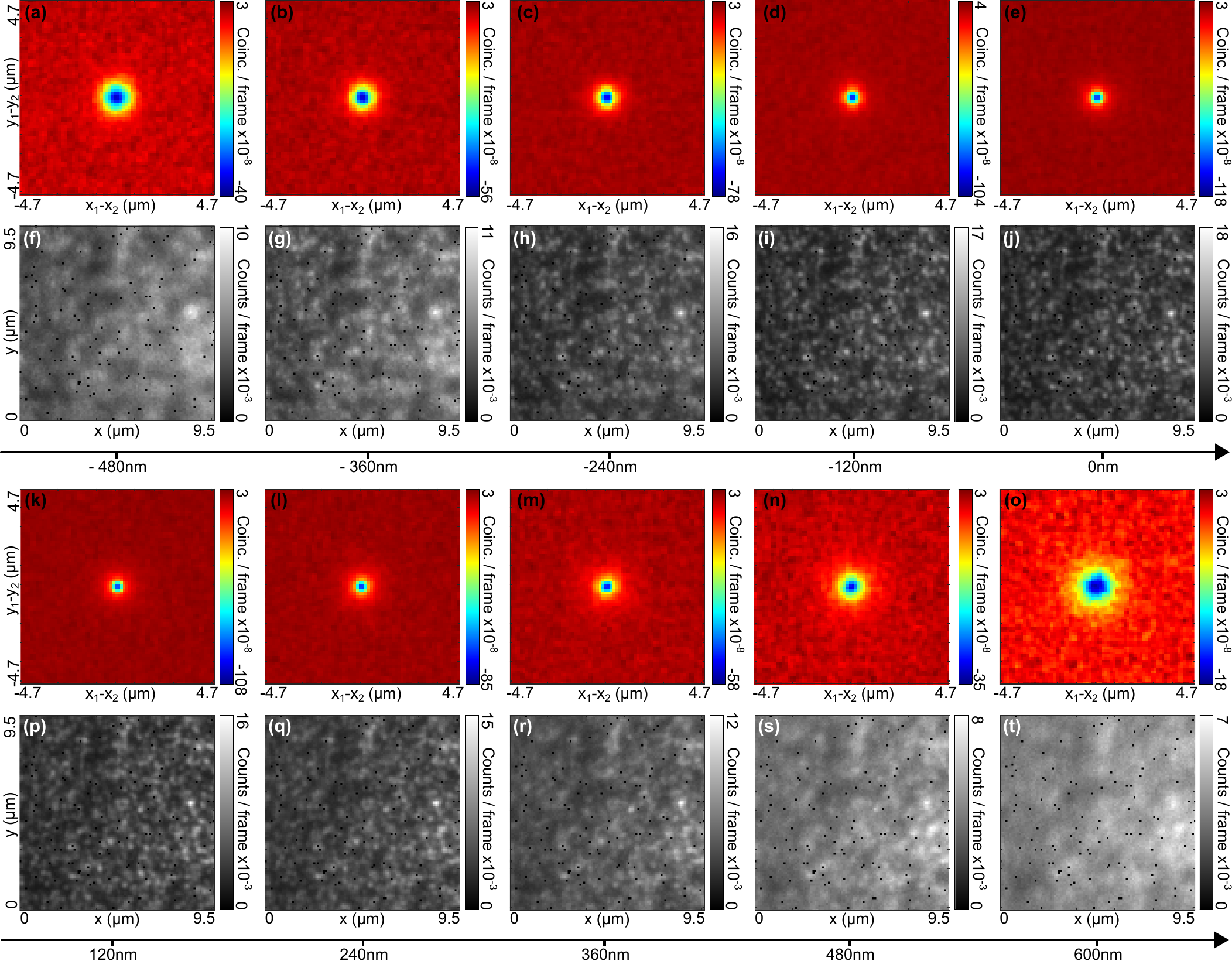}
\caption{\label{fig:sm_results_def} \textbf{Complete dataset for the experimental results for defocused aberrations.} 
Panels \textbf{(a-e)} and \textbf{(k-o)} display the resulting spatial covariance images ($C$), and \textbf{(f-j)} and \textbf{(p-t)}  show the corresponding intensity images ($I$),  at various axial positions. To induce defocus, the sample is translated axially across the focal plane.}
\end{figure*}

\begin{figure*}[t]
\centering
\includegraphics[width=1\linewidth]{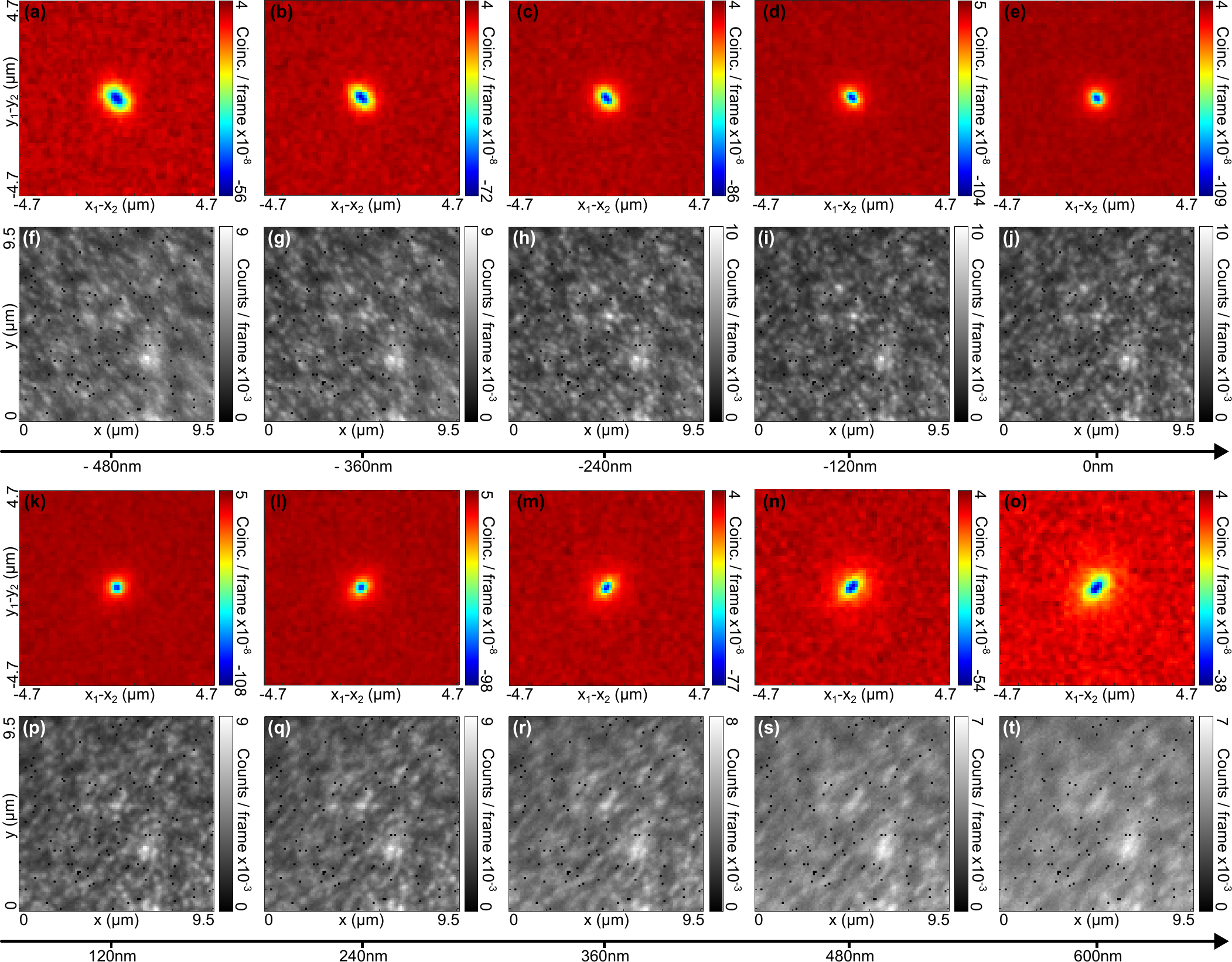}
\caption{\label{fig:sm_results_cyl_lens} \textbf{Complete dataset for the experimental results for defocused astigmatism aberrations.} 
Panels \textbf{(a-e)} and \textbf{(k-o)} display the resulting spatial covariance images ($C$), and \textbf{(f-j)} and \textbf{(p-t)}  show the corresponding intensity images ($I$), at various axial positions. To induce astigmatism, a cylindrical lens is introduced into the imaging path, rendering the PSF elliptical, and then the sample is translated along the axial direction.}
\end{figure*}

\Cref{fig:sm_results_def} and \Cref{fig:sm_results_cyl_lens} present additional data corresponding to the defocus and astigmatism aberration results shown in the main text. In both cases, the progressive broadening of the covariance function is clearly observed as the imaging plane moves away from the focal plane. For astigmatism, the covariance function also exhibits a change in elliptical orientation across the focal plane, consistent with the characteristic behavior of astigmatic aberrations. Video visualization of both cases can be observed in Visualization 1 (defocus) and Visualization 2 (astigmatism).

\begin{figure*}[t]
\centering
\includegraphics[width=1\linewidth]{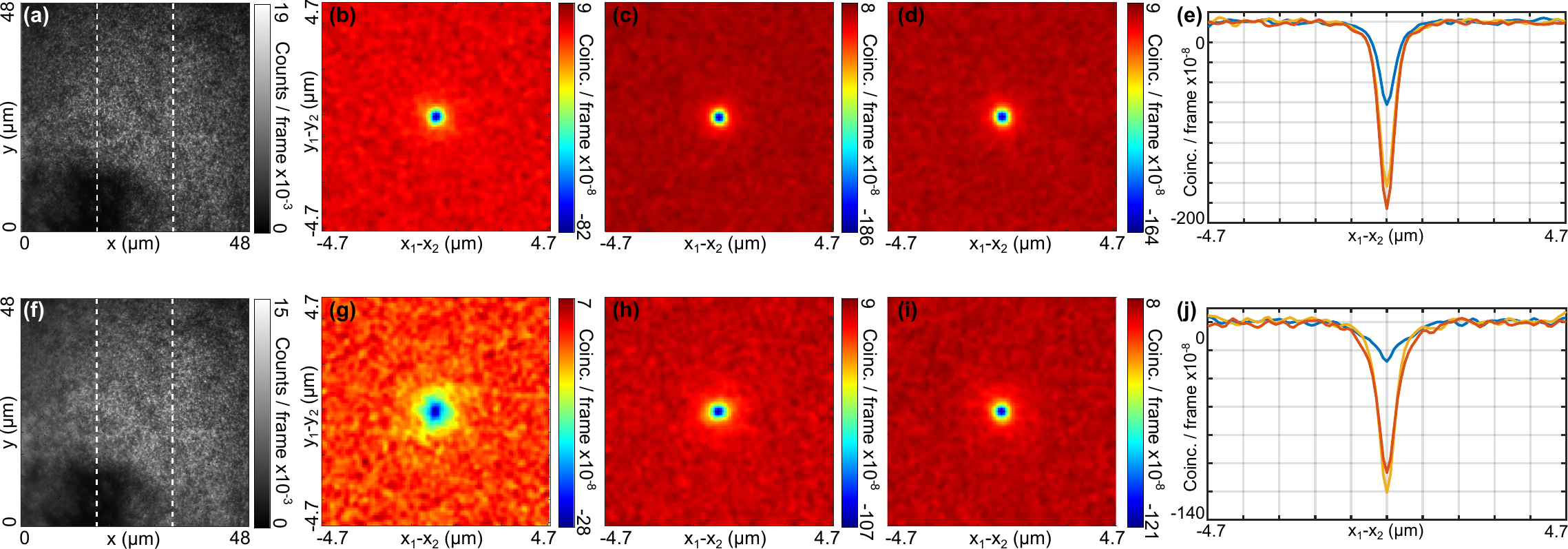}
\caption{\label{fig:sm_results_shift_var3x3} \textbf{Experimental results for a shift-variant aberration with defocus.}  The sample was defocused by $360$nm between configurations (a-e) and (f-j). The sample was tilted along a transverse axis such that the different parts of the sample experienced varying defocus, creating a shift-variant PSF. \textbf{(a,f)} Average intensity measurement within the full field of view (FOV). To reveal the shift-varying PSF, we calculate the spatial covariance $C(\vec{r})$ independently for three segments of the FOV, denoted by dashed lines in (a,f). \textbf{(b-d)} and  \textbf{(g-i)} Spatial covariance images for the left, center, and right segments. \textbf{(e,i)} 1D cross-sections at $y_1-y_2=0$ of the left (blue), center (red), and right (yellow) covariance images. 
}
\end{figure*}

\Cref{fig:sm_results_shift_var3x3} presents additional data illustrating the spatially varying PSF across different segments of the field of view. The same tilted sample as in the main text was used. Between configurations (a–e) and (f–j), the sample was axially displaced by 360 nm. The effect of defocus is clearly visible across the three field-of-view segments: the correlation function in the left segment broadens substantially from (b) to (g), while the plane of best focus shifts from the center segment in (c) toward the right segment in (i). This shift is reflected in the one-dimensional cross-sections in (j), where the right-segment profile (yellow) becomes narrower. Video visualization of this case can be observed in Visualization 3.

\begin{figure*}[t]
\centering
\includegraphics[width=1\linewidth]{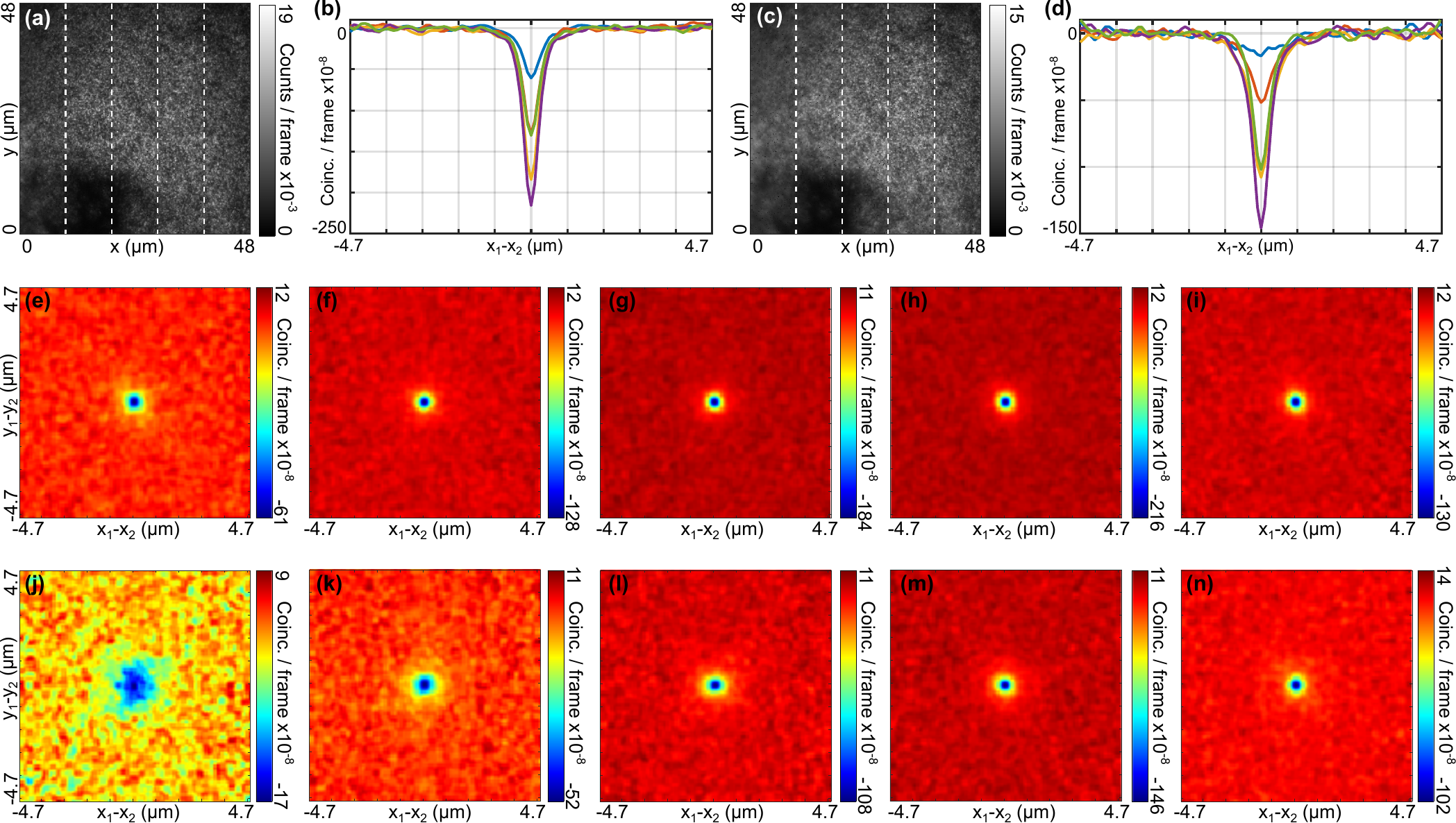}
\caption{\label{fig:sm_results_shift_var_5part} \textbf{Illustration of shift-variant aberration with a smaller patch size.} We use the same dataset and repeat the analysis shown in \Cref{fig:sm_results_shift_var3x3}, but divide the field of view (FOV) into five segments instead of three. \textbf{(a,c)} Average intensity measurement within the full field of view (FOV) for the two configurations. The spatial covariance $C(\vec{r})$ is calculated independently for five segments of the FOV, denoted by dashed lines in (a,c). \textbf{(e-i)} and  \textbf{(j-n)} Spatial covariance images for the five segments. \textbf{(b,d)} 1D cross-sections of the covariance images along $y_1-y_2=0$, corresponding to the five segments from left to right: blue, red, yellow, purple, and green.
}
\end{figure*}

\Cref{fig:sm_results_shift_var_5part} repeats the analysis for shift-variance PSF using a smaller patch size, with the field of view divided into five segments instead of three. Despite the reduced patch size, the spatial variation of the PSF across the field of view remains clearly observable. The smaller patches also lead to an increase in noise (see \Cref{fig:sm_results_shift_var_5part}(j)), however, it remains well-above the noise floor. Video visualization of this case can be observed in Visualization 4.

\begin{figure*}[t]
\centering
\includegraphics[width=1\linewidth]{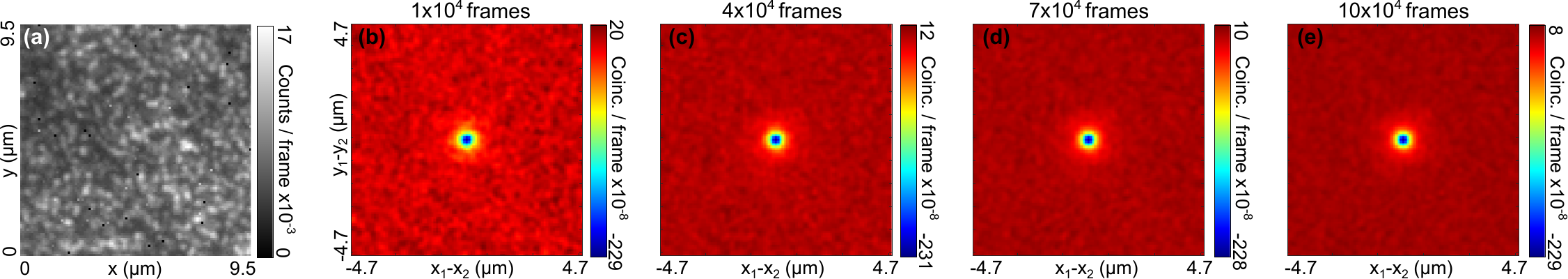}
\caption{\label{fig:sm_results_power5x} \textbf{Covariance measurement with lesser number of frames.} We increased the laser power to \SI{13}{\milli\watt} (x5 comparing to the rest of the examples) and analyzed the effect of number of frames on the covariance measurement. \textbf{(a)} Average intensity measurement. \textbf{(b-e)} Spatial covariance images for increasing number of frames. While the background noise (the maxima of the colorbar) gets lower with increasing number of frames, the covariance signal (minima of the colorbar) remains practically constant, and the overall PSF autocorrelation form is consistent.}
\end{figure*}

In \Cref{fig:sm_results_power5x}, the laser power was increased to \SI{13}{\milli\watt} (x5 comparing to the rest of the examples), and we demonstrate that the covariance signal remains robust even when only $10^4$ frames are acquired (comparing to $2\cdot10^5$ in all of the other experiments), corresponding to an exposure time of \SI{100}{\milli\second} in the maximal frame rate of the SPAD512 (in our experiments the exposure time was \SI{140}{\milli\second} due to the limitation of the laser repetition rate). Across panels (b–e), the covariance signal, represented by the minimum value of the color bar, remains nearly unchanged despite the different numbers of frames used. In contrast, the background noise, represented by the maximum value of the color bar, decreases with increasing frame number, as expected. These results indicate that the covariance-based measurement retains a strong signal even for relatively short acquisition times. The power increase results in some bleaching, however, most of it can be avoided with better shutter synchronization (as the laser shutter is open much longer than required). This example shows that with some further technical improvements, closed-loop auto-focus and AO can already be implemented in the current setup with near to real-time operation. 

\bibliography{bib}